\documentclass[prl,twocolumn, amsmath,amssymb,aps,superscriptaddress]{revtex4-2}
\usepackage{amsfonts}
\usepackage{stmaryrd}
\usepackage{bbm}
\usepackage{mathrsfs}
\usepackage{tipa}
\usepackage{amssymb}
\usepackage{txfonts}
\usepackage{graphicx}
\usepackage{dcolumn}
\usepackage{epstopdf}
\usepackage[colorlinks,linkcolor=blue,urlcolor=blue,citecolor=blue]{hyperref}
\usepackage{multirow}
\usepackage{subfigure}
\usepackage{url}
\usepackage{upgreek}
\usepackage{booktabs}
\usepackage[utf8]{inputenc}
\usepackage[english]{babel}
\usepackage[normalem]{ulem}

\begin{document}
	
\newcommand*{\Tc}{$T_c$ }
\newcommand*{\kFl}{$k_Fl$ }

\preprint{APS/123-QED}
	
\title{Anomalous terahertz nonlinearity in disordered s-wave superconductor close to the superconductor-insulator transition}% Force line breaks with \\

\author{Hao Wang}
\affiliation{International Center for Quantum Materials, School of Physics, Peking University, Beijing 100871, China}

\author{Jiayu Yuan}
\affiliation{International Center for Quantum Materials, School of Physics, Peking University, Beijing 100871, China}

\author{Hongkai Shi}
\affiliation{School of Electronic Science and Engineering, Nanjing University, Nanjing 210093, China}

\author{Haojie Li}
\affiliation{School of Electronic Science and Engineering, Nanjing University, Nanjing 210093, China}

\author{Xiaoqing Jia}
\affiliation{School of Electronic Science and Engineering, Nanjing University, Nanjing 210093, China}

\author{Xiaohui Song}
\affiliation{Beijing National Laboratory for Condensed Matter Physics, Institute of Physics, Chinese Academy of Sciences, Beijing 100190, China}

\author{Liyu Shi}
\affiliation{International Center for Quantum Materials, School of Physics, Peking University, Beijing 100871, China}

\author{Tianyi Wu}
\affiliation{International Center for Quantum Materials, School of Physics, Peking University, Beijing 100871, China}

\author{Li Yue}
\affiliation{Beijing Academy of Quantum Information Sciences, Beijing 100913, China}

\author{Yangmu Li}
\affiliation{Beijing National Laboratory for Condensed Matter Physics, Institute of Physics, Chinese Academy of Sciences, Beijing 100190, China}
\affiliation{School of Physical Sciences, University of Chinese Academy of Sciences, Beijing 100049, China}

\author{Kui Jin}
\affiliation{Beijing National Laboratory for Condensed Matter Physics, Institute of Physics, Chinese Academy of Sciences, Beijing 100190, China}
\affiliation{School of Physical Sciences, University of Chinese Academy of Sciences, Beijing 100049, China}
\affiliation{Songshan Lake Materials Laboratory, Dongguan, Guangdong 523808, China}

\author{Dong Wu}
\affiliation{Beijing Academy of Quantum Information Sciences, Beijing 100913, China}

\author{Jianlin Luo}
\affiliation{Beijing National Laboratory for Condensed Matter Physics, Institute of Physics, Chinese Academy of Sciences, Beijing 100190, China}

\author{Xinbo Wang}
\email{xinbowang@iphy.ac.cn}
\affiliation{Beijing National Laboratory for Condensed Matter Physics, Institute of Physics, Chinese Academy of Sciences, Beijing 100190, China}
	
\author{Tao Dong}
\email{taodong@pku.edu.cn}
\affiliation{International Center for Quantum Materials, School of Physics, Peking University, Beijing 100871, China}
	
\author{Nanlin Wang}
\email{nlwang@pku.edu.cn}
\affiliation{International Center for Quantum Materials, School of Physics, Peking University, Beijing 100871, China}
\affiliation{Beijing Academy of Quantum Information Sciences, Beijing 100913, China}
\affiliation{Collaborative Innovation Center of Quantum Matter, Beijing 100871, China}

\date{\today}% It is always \today, today,
%  but any date may be explicitly specified
	
\begin{abstract}
		 
Detection of the Higgs mode in superconductors using nonlinear terahertz spectroscopy is a key area of interest in condensed matter physics. We investigate the influence of disorder on the nonlinear terahertz response and the Higgs mode in NbN thin films with varying Ioffe-Regel parameters ($k_Fl$). In strongly disordered films near the superconductor-insulator transition (SIT), we observe an anomalous third-harmonic generation (THG) signal above $T_c$, which is absent in both cleaner superconducting and non-superconducting counterparts. The persistence of this normal-state THG signal in a high magnetic field excludes superconducting fluctuations as its origin. Below $T_c$, the THG intensity increases sharply, indicating a dominant contribution from the driven Higgs mode. The THG spectrum of the strongly disordered sample exhibits a broadened, multi-peak structure, which we attribute to quantum path interference between distinct channels involving unpaired electrons and Cooper pairs within emergent superconducting islands. Our findings not only demonstrate how disorder tunes the nonlinear terahertz response but also uncover a strong coupling between electrons responsible for normal-state THG and the superconducting Higgs mode below $T_c$ in strongly disordered samples.
		
\end{abstract}
	
%\keywords{Suggested keywords}%Use showkeys class option if keyword
%display desired
\maketitle

The spontaneous breaking of continuous symmetry creates collective excitations like amplitude and phase modes. In the case of superconductors, the phase mode is absorbed into the longitudinal component of the gauge field, raising its energy to the plasma frequency via the Anderson-Higgs mechanism\cite{anderson1}. This leads to the lowest energy collective excitation being the gapped amplitude or Higgs mode. Recent advances in intense terahertz (THz) spectroscopy have enabled the detection of the collective Higgs mode in a variety of superconductors \cite{Shimano2020Higgs,Shimano2013freeHiggs,Shimano2014NbNTHG,Shimano2019Nonlinear,Kaiser2020Phaseresolved, Kaiser2023Fano, Chu2023Dynamicala, ZheWang2021MgB2THG,ZXWang2022NbNHiggs, Shimano2017NbNTHG, MgB2_THG_H_field,Shimano2018Higgs,JGWang2021Higgs_BaFeCoAs,NLWang2024YBCOTHG,JGWang2025Higgs_echo}. It is generally believed that the disorder plays a crucial role in identifying the Higgs collective mode in nonlinear terahertz spectroscopy. At modest (nonmagnetic) disorder levels, while the {superconducting energy gap $\Delta$ remains robust against disorder \cite{Anderson1959Theory}, the THz-Higgs coupling strength can be significantly enhanced by disorder in the paramagnetic channel \cite{Benfatto2016CPvsHiggs,HideoAoki2016HiggsBeyondBCS,Silaev2019Nonlinear,Shimano2019Nonlinear,PhysRevResearch.2.043029,PhysRevB.104.134504,PhysRevB.106.144509}, consequently resulting in the Higgs-mode contribution to the THz nonlinear response substantially exceeding that from quasiparticles.

As disorder increases and the system enters the Anderson localization regime, enhanced fluctuations establish strongly disordered superconductors as a paradigm for studying quantum phase transitions\cite{QPT_PRL_Fisher,QPT_Kapitulnik,QPT_review_NP,QPT_review_JianWang}. A key feature is the emergence of superconducting granularity: islands with large pairing amplitudes are separated by regions of strongly reduced amplitude \cite{Larkin2001Disorder, Trivedi1998Role, Ioffe2011Localization, Lee1985Localized, Sanquer2008DisorderInduced, Avishai2007DisorderIsland}. Phase coherence is maintained within each island, while inter-island coherence weakens due to intensified phase fluctuations under strong disorder. A natural question arising in this context concerns the fate of the Higgs mode and its nonlinear response in such an inhomogeneous superconductor. Earlier theoretical studies have partially addressed the first part of this question, predicting that the Higgs mode softens but remains sufficiently well-defined near the SIT \cite{theory_Higggs_Linear_response, Arovas2011Visibility, Arovas2013Dynamics, the_fate_of_Higgs_2013}. Subsequent linear THz spectroscopy have observed features consistent with this predicted Higgs mode in disordered NbN films \cite{Dressel2015Higgs}, though this interpretation remains debated \cite{Armitage2016NbNTDS}. However, the nonlinear response of the Higgs mode near the SIT has not been experimentally investigated. 

The second feature associated with emergent inhomogeneous superconductivity in the strongly disordered regime is the appearance of a pseudogap above $T_c$. Previous scanning tunneling spectroscopy studies \cite{Pratap2012PhaseDiagram,Raychaudhuri2013NbNSTS} have revealed that in strongly disordered NbN near the SIT, the single-particle gap $\Delta$ persists above $T_c$ and forms a pseudogap. Recent nonlinear THz studies on disordered NbN \cite{Armitage2022NbNnonlinear} have shown that, in the same strongly disordered regime, the nonlinear response extends up to nearly ten times $T_c$, an effect attributed to superconducting fluctuations. This hypothesis, however, requires experimental verification. 

\begin{figure*}
	\centering
	\includegraphics[width=7in]{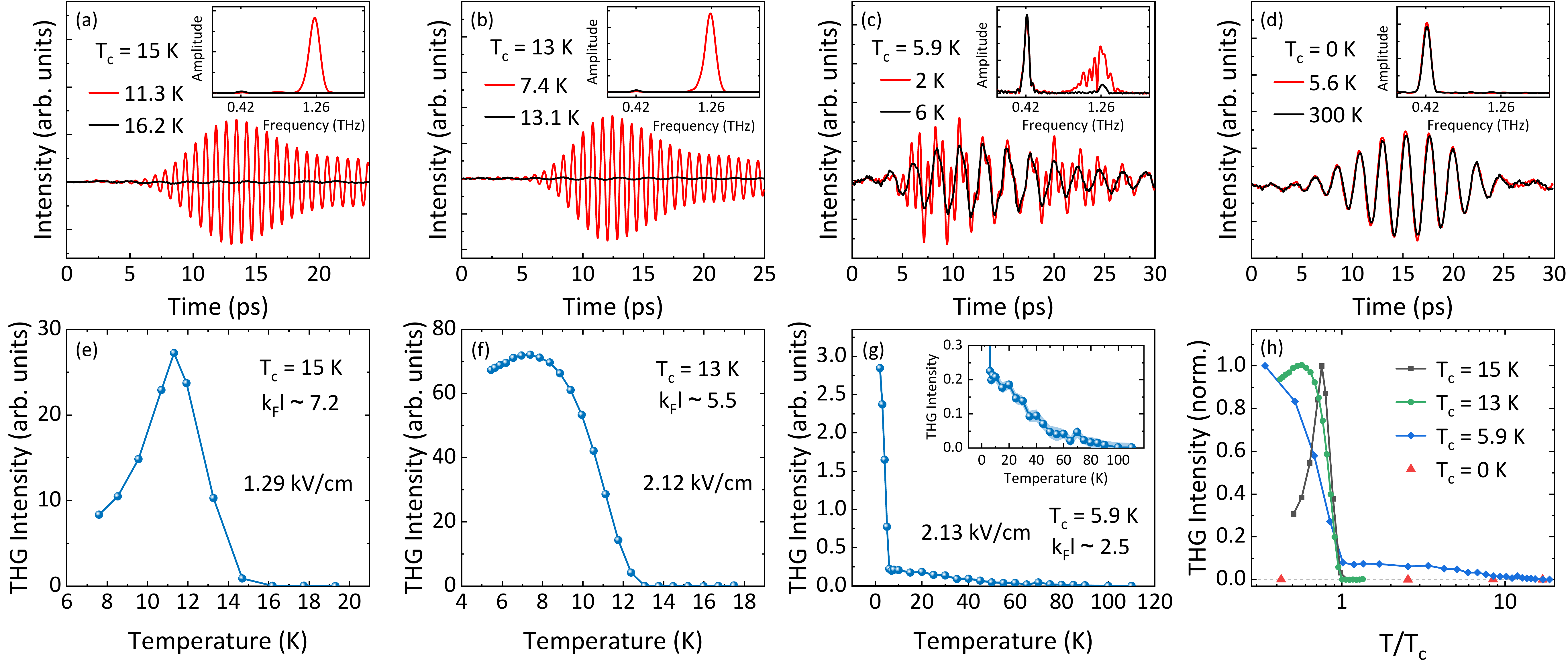}	
	\caption{THG response in NbN films with varying disorder levels. (a–d) Time-domain transmitted THz pulses for disordered NbN with $T_c$ values of (a) 15~K, (b) 13~K, (c) 5.9~K, and (d) 0~K under 0.42~THz driving. Insets show the corresponding Fourier transform spectra. A 1.26~THz bandpass filter was applied post-sample to attenuate the fundamental frequency. (e–g) Temperature-dependent THG intensity for samples with $T_c$ = (e) 15~K, (f) 13~K, and (g) 5.9~K. The estimated pump electric field strengths inside the samples (taking into account the Fabry–Pérot effect) are labeled in the figures, which are about one order smaller than the field strength before the sample due to the low transmission coefficient. The inset in (g) highlights the THG intensity above $T_c$. THG intensity is defined as the spectral integral of the $3\omega$ peak. Shaded regions represent error bounds (noise floors of frequency spectra). Error bars for $T_c$ = 13~K and 15~K are negligible relative to line widths. (h) Normalized THG intensity versus temperature. For the non-superconducting sample ($k_Fl < 1$), $T_c$ = 5.9~K is used for $T/T_c$ normalization.}
	
	\label{fig:NbN THG}
\end{figure*}

In this work, to address above issues, we conduct systematic THz THG studies on NbN thin films with four distinct disorder levels. In the strongly disordered sample ($k_Fl \sim 2.5$), near the SIT, two distinct features are observed. Above $T_c$, a weak yet detectable THG signal persists. Below $T_c$, normal-state THG interacts with inhomogeneous superconductivity-induced signals, producing multiple peaks in the THG spectra. By applying a magnetic field to suppress global superconducting coherence and comparing with the $k_Fl \sim 7.2$ sample, we attribute the multi-peak feature to mesoscopic superconducting inhomogeneity and propose that the normal-state nonlinearity originates from a mechanism independent of superconductivity.

We drive the superconducting films with $f$=0.42 THz pulse, the third-order harmonic $3f$=1.26 THz radiations are recorded. The experimental details are provided in the Supplemental Material (SM). Fig~\ref{fig:NbN THG}(a–d) shows the time-domain waveforms of transmitted THz pulses for NbN samples at different disorder levels, with the corresponding fast Fourier transform (FFT) spectra shown in the insets. Each panel presents two waveforms: one measured at the temperature of maximum THG intensity and another just above $T_c$. Panels (e–g) display the temperature-dependent THG intensity for the samples in Fig.~\ref{fig:NbN THG}(a–c). For moderately disordered samples with $T_c = 15$ K and 13 K, the THG intensity is significantly enhanced in the superconducting state. Resonant features appear when the superconducting gap equals twice the driving frequency, i.e., $2\Delta(T) = 2\omega$ \cite{Shimano2014NbNTHG,Aoki2015HiggsTheory,Benfatto2016CPvsHiggs,ZXWang2022NbNHiggs}. Above $T_c$, the THG intensity vanishes completely, consistent with prior studies \cite{Shimano2013freeHiggs,Shimano2014NbNTHG,ZXWang2022NbNHiggs}.

As the disorder level increases to $k_Fl \sim 2.5$ (see SM for the estimation of $k_Fl$), the NbN film with $T_c = 5.9$ K exhibits distinct behavior. At 2 K, the time-domain waveform shows separated wavelets, and its FFT spectrum contains multiple peaks with a low-frequency tail. This contrasts with the $k_Fl \sim 7.2$ and $k_Fl \sim 5.5$ samples, where the THG spectra display single peaks. We will discuss the multi-peak feature later. The THG intensity drops sharply at $T_c$ but persists at a finite value, gradually decreasing up to 100 K—again, contrasting with moderately disordered samples. Above $T_c$, the THG spectrum reverts to a single peak centered at three times the driving frequency. Note that the normal-state nonlinearity of the $k_Fl \sim 2.5$ sample is significantly weaker than in its superconducting state, cuprate superconductors, and linearly dispersive materials. Upon further increasing the disorder to completely suppress superconductivity ($k_Fl < 1$, Fig.~\ref{fig:NbN THG}(d)), THG becomes undetectable across the accessible temperature range. Fig~\ref{fig:NbN THG}(h) summarizes the normalized THG as a function of $T/T_c$ for all measured samples, showing the extension of the THG signal up to 10$T_c$ for the $k_Fl \sim 2.5$ sample. Furthermore, the consistent enhancement of THG intensity at $T_c$ across different disordered samples suggests that the Higgs mode still dominates the THG process in the strongly disordered sample. Nonetheless, a THz pump-probe study is needed to resolve the free oscillation and life time of the Higgs mode in the $k_Fl \sim 2.5$ sample. 

Given the correlation between normal-state nonlinearity and low-temperature superconductivity, one may attribute the observed effects to local superconducting fluctuations that persist well above the global $T_c$, with their nonlinearities governing the response. Indeed, previous THz nonlinear studies reported similar normal-state nonlinearity in disordered NbN \cite{Armitage2022NbNnonlinear}, exhibiting a dome-shaped dependence on disorder within the range $1 < k_Fl < 4$. The authors attributed this nonlinearity to superconducting fluctuations. To test this hypothesis, we apply a magnetic field to tune the superconducting parameters \cite{MgB2_THG_H_field,Wang2024Tabletop}; if the normal-state THG arises from superconducting fluctuations, it should be suppressed with increasing magnetic field.

\begin{figure}
	\centering
	\includegraphics[width=3.3in]{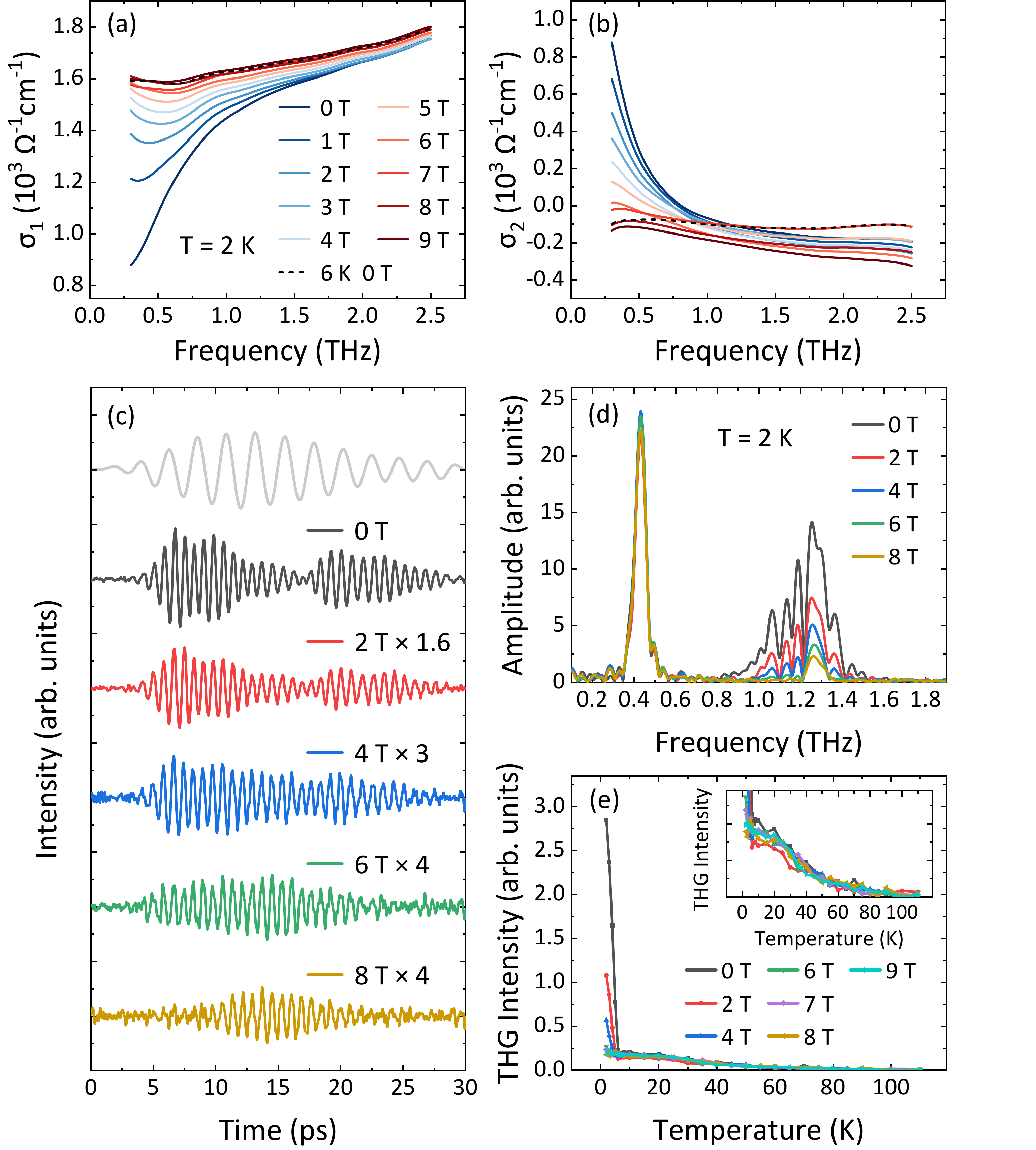}		
\caption{Magnetic field evolution of the linear and nonlinear THz response for the $T_c = 5.9$~K sample. (a) Real part, $\sigma_1(\omega)$, and (b) imaginary part, $\sigma_2(\omega)$, of optical conductivity at various magnetic fields, measured at 2~K. The magnetic field is applied perpendicular to the sample surface. The dashed black line represents the reference measurement at 6~K under zero magnetic field. (c) Time-domain waveforms of the THG signal measured at 2~K under different magnetic fields. The light gray curve at the top shows the fundamental waveform. Both THG and fundamental components are extracted from the transmitted waveforms (see SM) by digital filtering: THG waveforms are obtained using a high-pass filter with a cutoff frequency of 0.75~THz, while the fundamental waveform is obtained using a low-pass filter with the same cutoff frequency. (d) FFT of transmitted THz pulses measured at 2~K under different magnetic fields. A multi-peak feature appears in the superconducting state. (e) Temperature dependence of THG intensity at different magnetic field strengths. The inset shows a magnified view of the normal-state signal.}	
	\label{fig:NbN6K}
\end{figure}

Before performing THG measurements under magnetic fields, we characterized the field-dependent optical conductivity of the $T_c = 5.9$~K NbN sample using magneto-terahertz time-domain spectroscopy. Fig~\ref{fig:NbN6K}(a) and \ref{fig:NbN6K}(b) show the field dependence of the real part, $ \sigma_1(\omega) $, and imaginary part, $ \sigma_2(\omega) $, of the optical conductivity at 2~K. As the magnetic field increases, the characteristic dip in $ \sigma_1(\omega) $ and the $ 1/\omega $ feature in $ \sigma_2(\omega) $ gradually diminish, eventually returning to a normal-state-like response above the critical field of 7~T.

We next examine THG measurements under magnetic fields perpendicular to the sample surface. Fig~\ref{fig:NbN6K}(c) and (d) show the time-domain waveforms and corresponding FFT spectra of the magnetic-field-dependent THG at 2~K. Below $ T_c $, increasing magnetic field progressively suppresses superconductivity, reducing the THG amplitude. In contrast, Fig.~\ref{fig:NbN6K}(e) and its inset show no detectable change in THG intensity above $ T_c $, even at magnetic fields up to 9~T. If superconducting fluctuations were responsible for normal-state nonlinearity, such high fields would reduce local pairing amplitudes, thereby diminishing the THG intensity. The magnetic field insensitivity of THG intensity indicates that normal-state THG originates from disorder-induced material properties rather than superconducting fluctuations. The observation of THG signals up to room temperature under 0.7 THz driving with elevated electric fields (see Fig.~\ref{0.7} in SM) further supports this conclusion. At 2~K, the 7~T magnetic field fully suppresses global phase coherence while preserving localized Cooper pairs in islands \cite{Avishai2007DisorderIsland}, yet the THG magnitude remains unchanged. This suggests that localized Cooper pairs in s-wave superconductors do not generate a significant THz nonlinear response. 

%\textcolor{red}{\sout{Whether mobile Cooper pairs without macroscopic coherence—such as those in ultrathin NbSe$_2$ \cite{ultrathin_NbSe2_Higgs} and nano-patterned cuprate superconducting films \cite{YBCO_nao_pattern}—produce significant THG remains an open question. }}

\begin{figure}
	\centering
	\includegraphics[width=0.5\textwidth]{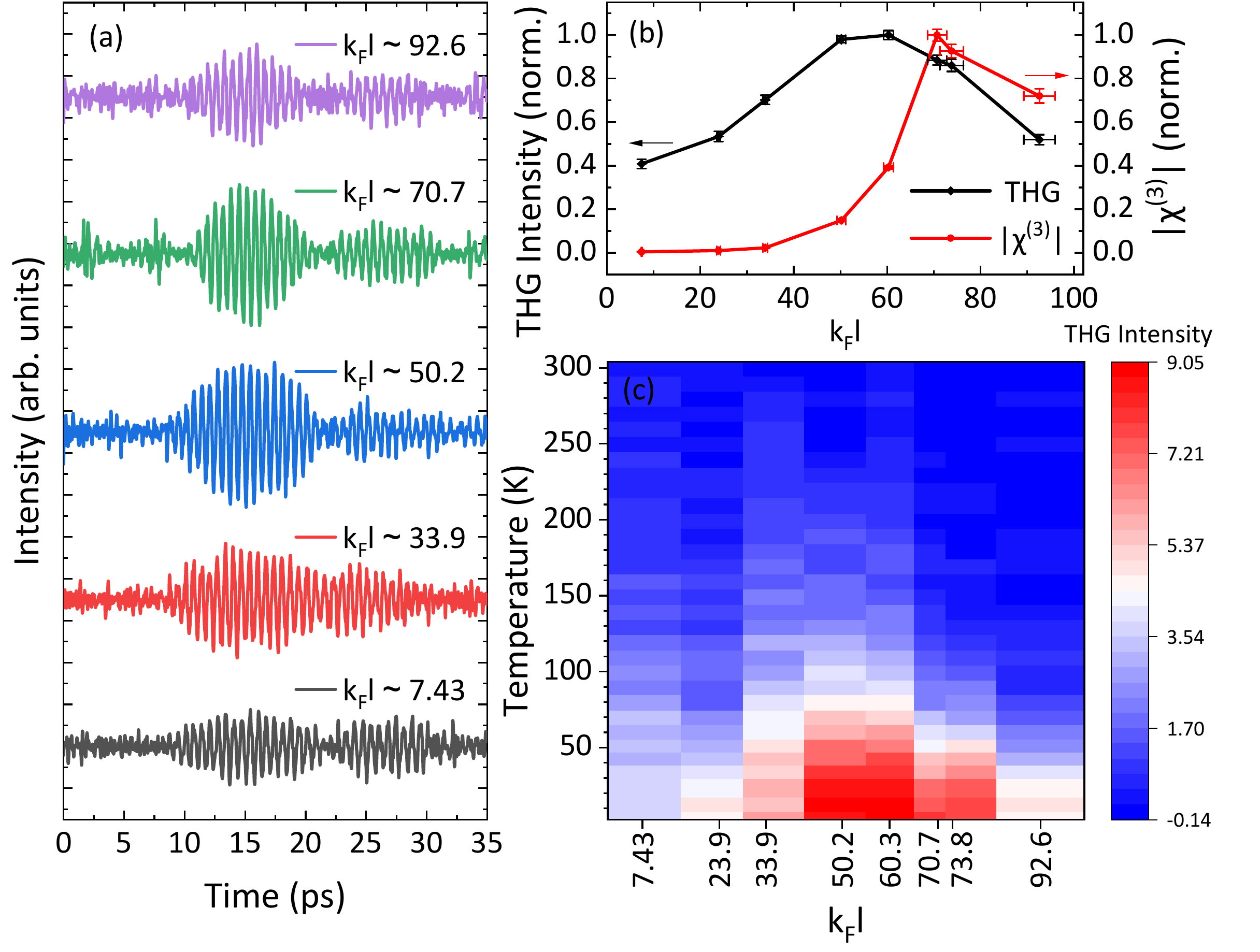}	
\caption{THG signal of disordered Au films driven by 0.5~THz. (a) 1.2~THz high-pass filtered time-domain waveforms of transmitted THz pulses for different $k_Fl$ values measured at 5~K. (b) $k_Fl$ dependence of normalized THG intensity and $|\chi^{(3)}|$ measured at 5~K. $|\chi^{(3)}|$ is estimated by $I_{3\omega}/I_\omega^3$, where the intensity of the fundamental pulse, $I_\omega$, was measured at 5~K without a 3$\omega$ bandpass filter. (c) Heat map showing the temperature and $k_Fl$ dependence of THG intensity.}
	\label{fig:Au THG}
\end{figure}

To further validate our conclusions and examine disorder-enhanced THG without superconductivity contributions, we performed comparative THG measurements on 10-nm-thick nonsuperconducting Au films with disorder levels ranging from $ k_F l = 7.4 $ to 92.6 (see SM for sample preparation). Fig~\ref{fig:Au THG} (a) presents the time-domain waveforms of THG signals at 5 K for selected disorder levels, after digital high-pass filtering at 1.2 THz. Although the THG signal from Au films is weak, it remains distinguishable above the noise floor. The second wave packet at about 11~ps after the first one originates from the substrate echo. Fig~\ref{fig:Au THG}(b) shows the normalized THG intensity at 5 K as a function of $ k_F l $, revealing a dome-shaped disorder dependence that peaks near $ k_F l \sim 60 $. After accounting for the effective field in the Au thin films, the normalized third-order susceptibility $ |\chi^{(3)}| \propto I_{3\omega} / I_\omega^3 $, where $ I_\omega $ is the fundamental intensity measured without 3$ \omega $-bandpass filtering, exhibits a similar trend, peaking near $ k_F l \sim 70 $. Fig~\ref{fig:Au THG}(c) illustrates the temperature and $ k_F l $ dependence of THG intensity via a heat map. The consistent dome-shaped nonlinearity in the normal state of both NbN and Au suggests that this disorder effect is universal across materials. Note that the Au thin film has much better conductivity, the dome-shaped nonlinearity appears at higher $ k_F l$.

Having ruled out superconductivity as the source of the normal-state nonlinearity in disordered NbN, we now examine its origin. Given the featureless normal state of NbN and the absence of any reported symmetry-breaking orders, the nonlinearity likely arises from disorder-induced modifications to the electronic structure. At THz frequencies, photons primarily excite electrons near the Fermi level; thus, THG in metals with a high Fermi energy is typically governed by intraband dynamics. While materials with nonparabolic bands are known to exhibit strong nonlinear responses \cite{Nakajima2024Complex, Pearson1966Theory, Wang2020Nonperturbative}, the negligible normal-state THG in our least disordered sample suggests a nearly parabolic dispersion in clean NbN. Disorder causes distortions in the periodic lattice potentials, which in turn distorts band structures and Fermi surfaces \cite{Aoki1984Electronic, Kasinathan2006Destruction}, potentially enhancing anharmonicity near the Fermi level and amplifying nonlinearity. Although this band anharmonicity remains experimentally unverified, our measurements at 0.7 THz under elevated electric fields reveal a persistent THG signal up to room temperature (see Fig.~\ref{0.7} in SM), suggesting that it provides a small but finite contribution.

Alternatively, nonlinearity can originate from energy-dependent scattering \cite{Wang2024CaRuO3} or non-Fermi-liquid behavior \cite{Volkov2025strangemetal}, though these are likely minor contributions since disorder predominantly enhances elastic scattering. A more significant mechanism arises from the disparity between energy and momentum relaxation times. Third-order nonlinearity is enhanced when the energy relaxation time, $\tau_E$, exceeds the momentum relaxation time, $\tau_M$, with the magnitude scaling as $\tau_E/\tau_M$ \cite{Wolff1982Differencefrequency}. Here, $\tau_M$ is the Drude scattering time \cite{Armitage2025Planckian}, while $\tau_E$ characterizes energy transfer from electrons to the lattice via electron-phonon coupling. Disorder drastically reduces $\tau_M$ but only weakly affects $\tau_E$, which is largely disorder-independent \cite{Goltsman2023Evidence}. This disparity enhances the THG signal with increasing disorder. Conversely, stronger disorder also intensifies electron localization and scattering, which suppress nonlinearity. The competition between these enhancement and suppression effects naturally explains the observed dome-shaped disorder dependence of the THG signal.

%The pseudogap is believed to correlate strongly with Cooper pairs; in highly disordered samples, \Tc is governed by the loss of global phase coherence, while the local pairing amplitude $|\Delta|$ remains finite due to phase-incoherent Cooper pairs, resulting in pseudogap behavior. However, experiments on disordered NbN reveal that the pseudogap persists up to approximately 7~K \cite{Pratap2012PhaseDiagram,Raychaudhuri2013stiffness}. Moreover, the pseudogap onset temperature cannot exceed \Tc of the cleanest samples, suggesting Cooper pairs cannot exist at the elevated temperatures where nonlinearity occurs.
\begin{figure}
	\centering
	\includegraphics[width=3.2in]{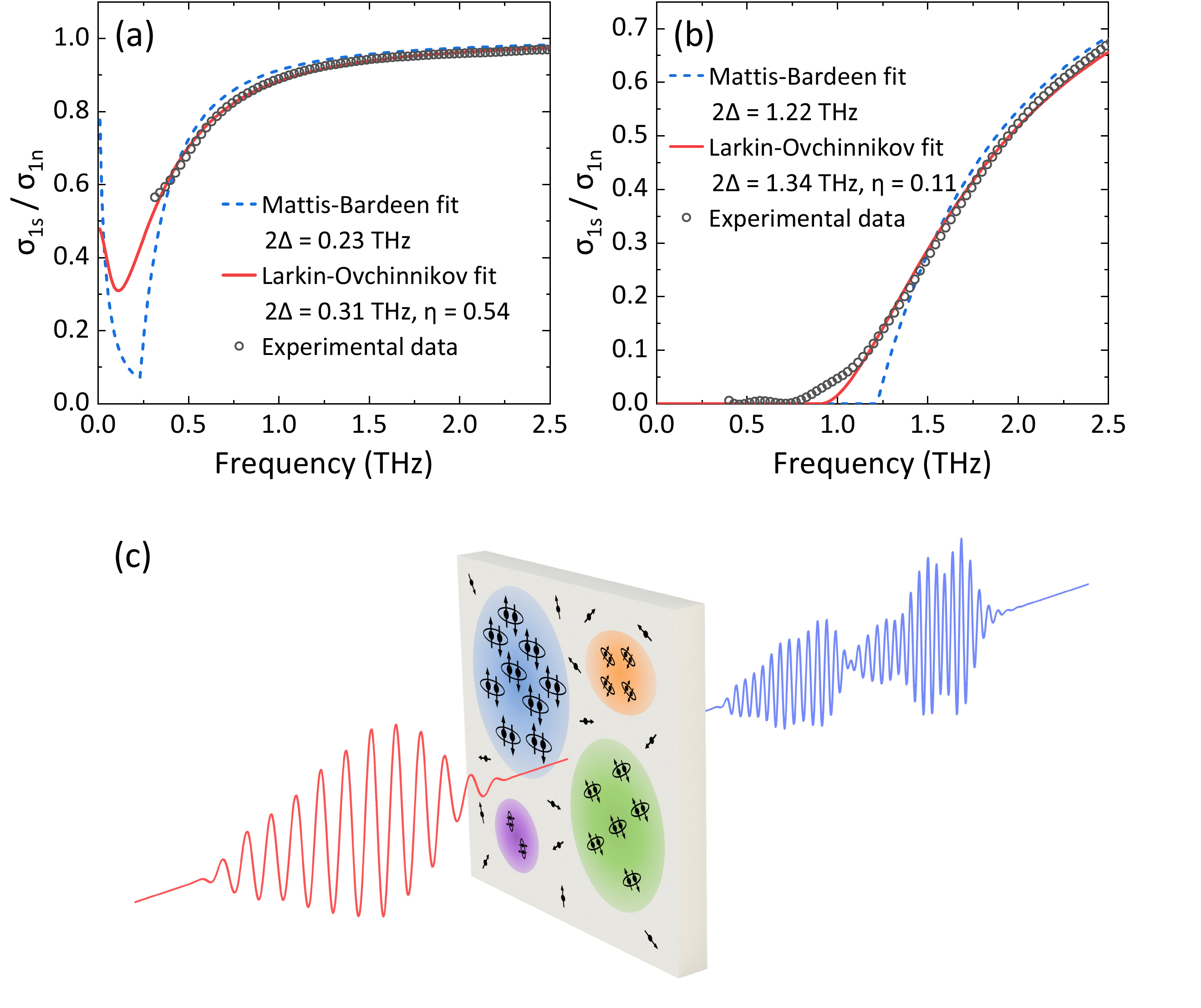}	
	\caption{Optical conductivity in the superconducting state. The normalized real part of the optical conductivity, $\sigma_1(\omega)/\sigma_1(N)$, is shown for (a) a sample with $k_Fl \sim 2.5$ and (b) a sample with $k_Fl \sim 5.5$, measured at 2 K. The blue dashed line and red solid line correspond to fits using the Mattis-Bardeen and Larkin-Ovchinnikov models, respectively. (c) Schematic of THG in a disordered sample, where superconducting islands are depicted as circles. The size and number of Cooper pairs represent the pairing amplitude, while their alignment indicates phase coherence.}		
	\label{fig:LOfit}
\end{figure}

%Beyond the normal-state THG, strong disorder gives rise to anomalous features in the superconducting state as well.
Regardless of its precise origin, normal-state nonlinearity may interfere with superconducting nonlinearity and produce anomalous nonlinear responses below $T_c$, similar to observations in cuprate superconductors \cite{Kaiser2020Phaseresolved,Kaiser2023Fano,NLWang2024YBCOTHG}. As shown in Fig.~\ref{fig:NbN6K}(c,d), a beating feature emerges in the time-domain waveforms, accompanied by interference peaks (multi-peak structure) in the frequency spectra. When global phase coherence of superconductivity is fully suppressed—either by a magnetic field exceeding 7~T or by heating above $T_c$—both the time-domain beating feature and the spectral multi-peak structure disappear abruptly. Supporting evidence from piecewise FFTs and wavelet transforms is provided in Fig.~\ref{piecewiseFFT}(a,b) and Fig.~\ref{wavelet} in the SM. These observations confirm that the phenomenon occurs exclusively in the superconducting state. Multiple reflection effects are ruled out, as their intensity is too weak to produce such pronounced phenomena (see SM for further analysis). Compared to the single THG spectral peaks observed in samples with $k_Fl \sim 7.2$ and $k_Fl \sim 5.5$, we attribute the interference peaks to mesoscopic superconducting inhomogeneity in strongly disordered superconductors.

In strongly disordered samples, a short elastic scattering length enhances localization, leading to mesoscopic inhomogeneity in the pairing potential $\Delta(\vec{r})$. To quantify this inhomogeneity for our sample with $k_F l \sim 2.5$, we analyze the optical conductivity. Fig~\ref{fig:LOfit}(a) shows the real part of the conductivity $\sigma_1(\omega)$. Below $T_c$, gap-opening features emerge in $\sigma_1(\omega)$ \cite{MattisBardeen1958MBtheory}, accompanied by a corresponding $1/\omega$ dependence in $\sigma_2(\omega)$. However, conventional Mattis-Bardeen theory \cite{MattisBardeen1958MBtheory} (blue dashed line in Fig.~\ref{fig:LOfit}(a)) fails to describe $\sigma_1(\omega)$ due to disorder-induced mesoscopic gap inhomogeneity. Instead, we use the Larkin-Ovchinnikov (LO) model \cite{Nam1967LOfit, Armitage2016NbNTDS, LO_model, Klapwijk2012LO} to characterize spatial variations in the superconducting gap. Key parameters of the LO model are the superconducting gap, $\Delta$, and the depairing strength, $\eta$, which directly reflect mesoscopic inhomogeneity in superconductivity. Details of the fitting procedure are provided in the SM. Using $2\Delta = 0.31$~THz and $\eta = 0.54$, our fit (red solid line in Fig.~\ref{fig:LOfit}(a)) captures the low-frequency absorption behavior of $\sigma_1$. A comparative fit for the sample with $k_F l \sim 5.5$ yields $\eta = 0.11$, indicating significantly reduced spatial variations. According to previous theoretical studies \cite{Avishai2007DisorderIsland}, this $\eta$ value suggests the presence of spatial variations in superconductivity, including inhomogeneities in the pairing amplitude and non-superconducting regions. Consequently, two distinct sources of THG coexist: coherent Cooper pairs and normal-state electrons, as illustrated in Fig.~\ref{fig:LOfit}(c). Within each superconducting island, THG from the Higgs mode couples with normal-state THG, causing a beating signal in the time domain waveform and energy splitting in the frequency domain. The strong coupling of THG signals from different sources is crucial, as evidenced by the broader split peaks compared to the THG spectra of its normal-state and the cleaner superconducting sample. Variations in pairing amplitude, phase, and coupling strength among islands generate distinct THG pathways. Interference between these pathways results in the observed complicated beating waveform in the time domain and the multi-peak structure in the frequency spectrum, serving as an alternative indicator of superconducting inhomogeneity. The low-frequency tail length and the number of interference peaks may reflect the spatial variations in superconductivity. Systematic measurements on strongly disordered samples ($1 < k_F l < 4$) and complementary theoretical studies are essential to establish quantitative relationships between interference features and mesoscopic superconducting inhomogeneity. Furthermore, extending the measurements to 2D THz scans will facilitate a complete understanding of how disorder influences the collective modes of the superconducting order parameter near the SIT.

In summary, we observe a weak yet detectable normal-state THG signal in strongly disordered NbN near the SIT, whereas no such signal is present in samples with either the highest or lowest levels of disorder. When superconductivity is suppressed by an applied magnetic field, the THG intensity of the normal-state nonlinearity remains unchanged, indicating that the observed normal-state nonlinearity in strongly disordered NbN arises from disorder-induced modifications to the band structure or scattering processes, rather than superconducting correlations. Below $T_c$, a sharp enhancement in THG intensity indicates that the Higgs mode remains the dominant contribution to the nonlinear response near the SIT. Furthermore, the appearance of interference peaks in the superconducting state of strongly disordered samples may serves as a probe of mesoscopic superconducting inhomogeneity. Our work provides key evidence to distinguish the origin of anomalous normal-state nonlinearity, advancing the understanding of disorder's role in the nonlinear response of superconductors and bearing relevance to studies of high-temperature superconductors.

We gratefully acknowledge Y. Wan for illuminating discussions. This work was supported by the National Natural Science Foundation of China (No.~12488201, 12250008, 62271242, 12134018, 12274439), the National Key Research and Development Program of China (2021YFA1400200, 2022YFA1403901, 2022YFA1403400, 2024YFA1611300), the Innovation Program for Quantum Science and Technology (No.~2021ZD0303401), and the Synergetic Extreme Condition User 
Facility (SECUF, https://cstr.cn/31123.02.SECUF). 

\bibliographystyle{apsrev4-2}	
\bibliography{citearticles}% Produces the bibliography via BibTeX.

\clearpage
\onecolumngrid

\renewcommand{\thefigure}{S\arabic{figure}}
\renewcommand{\thetable}{S\arabic{table}}
\setcounter{figure}{0}  

\setcounter{secnumdepth}{2}
\makeatletter
\renewcommand\section{\@startsection
	{section}{1}{0pt}%
	{5ex plus 1ex minus .2ex}%
	{2ex plus .2ex}%
	{\normalfont\bfseries\raggedright}} 
\makeatother

\begin{center}
	\textbf{\Large Supplemental material for "Anomalous terahertz nonlinearity in disordered s-wave superconductor close to the superconductor-insulator transition"}
\end{center}

\section{sample preparation}

The 50~nm thick NbN and 10~nm thick Au thin films were grown by magnetron sputtering on MgO single crystalline substrates with a thickness of 0.5~mm. Disorder in NbN was tuned by controlling the level of Nb vacancies in the lattice, which was achieved by changing the Nb/N ratio in the plasma \cite{Raychaudhuri2008NbN,Raychaudhuri2009NbN,NbN_growth}. The resistivity of the measured NbN films is shown in Fig.~\ref{resistivity}. The zero resistance temperatures of lower disordered samples are approximately 1 K higher than the temperatures at which THG disappears. In the main text, we use $T_c$ characterized by THG signal. For Au films, disorder was introduced by Ion Beam Milling System. Au films were bombarded by Ar$^+$ with low energy (200~eV) and flux ($4\times10^{19}$ ions$\cdot$s$^{-1}$m$^{-2}$) to introduce defects without changing the sample thickness. The disorder levels were tuned by the bombardment duration (10~$\sim$~40~s). We use Atomic Force Microscopy (AFM) to check the thickness and show the results in Fig.~\ref{AFM}. The thickness remains unchanged after being bombarded by Ar$^+$.

\begin{figure}[h]
	\centering
	\includegraphics[width=3in]{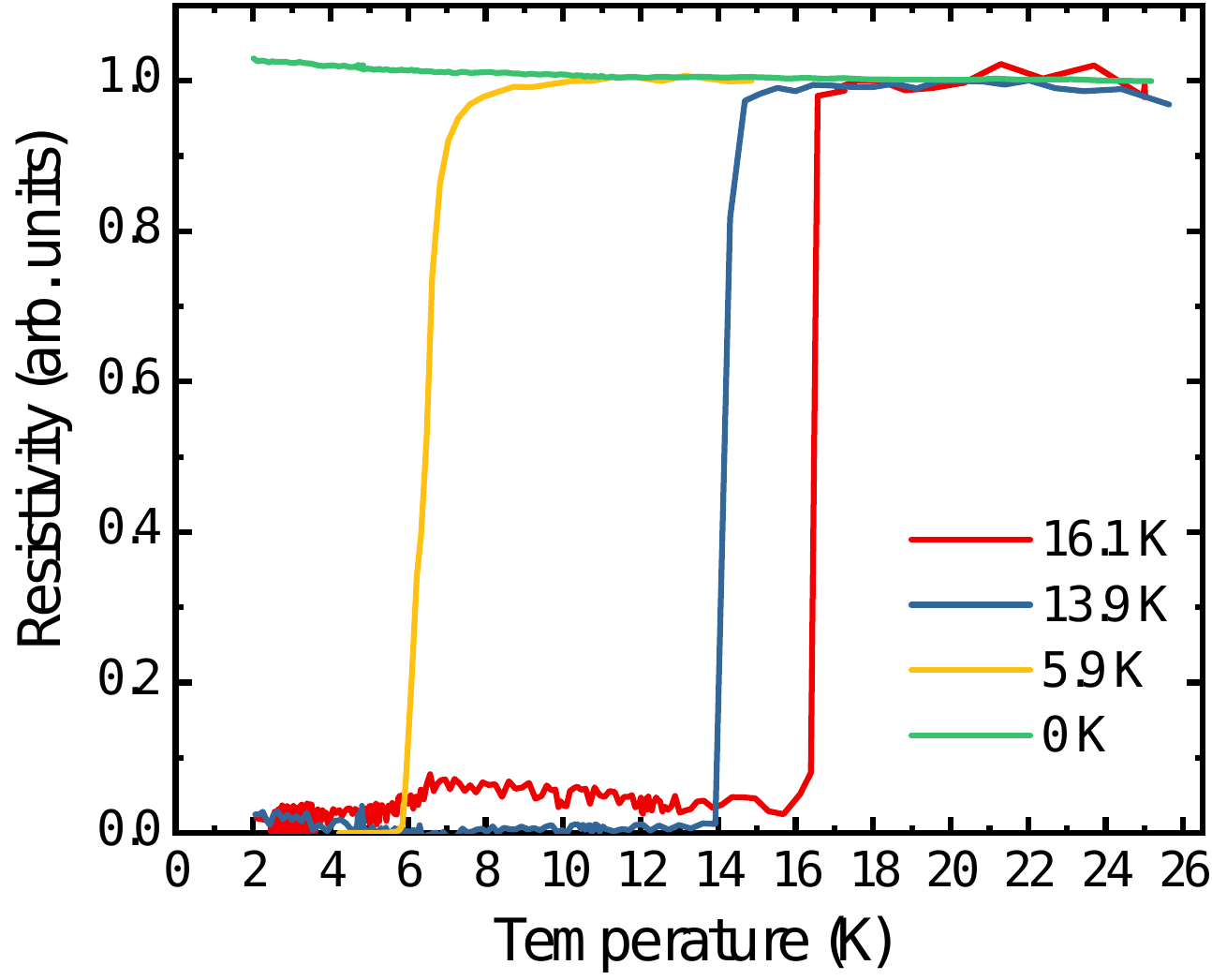}
	
	\caption{Resistivity of the four disordered NbN films.}
	
	\label{resistivity}
\end{figure}

\begin{figure}
	\centering
	\includegraphics[width=6.5in]{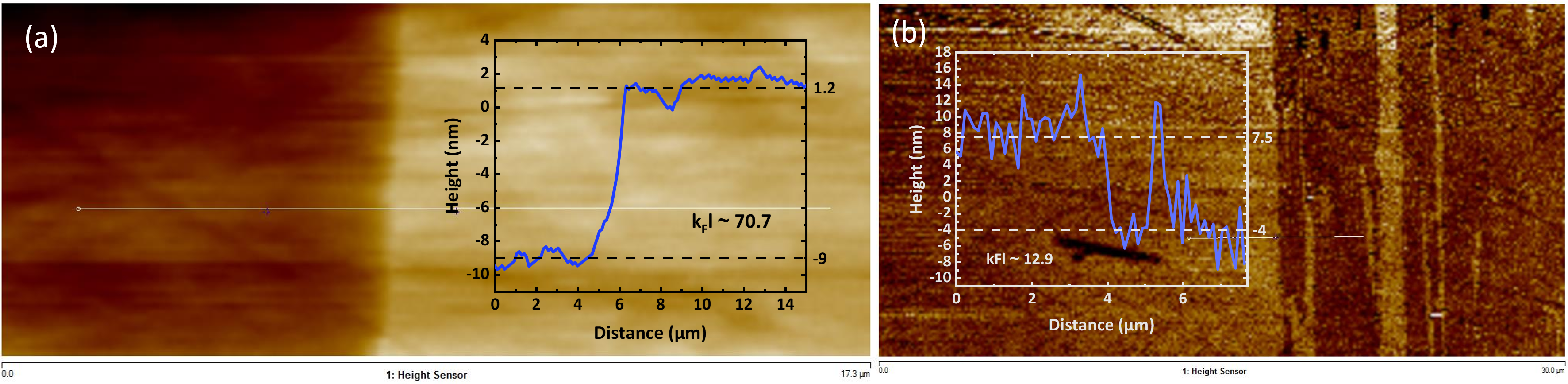}
	
	\caption{AFM measurement of Au films with (a) \kFl$\sim70.7$, (b) \kFl$\sim12.9$.}
	
	\label{AFM}
\end{figure}

\section{disorder level quantification}

To estimate disorder quantitatively, we used the Ioffe-Regel parameter \cite{Ioffe1960kFl} $k_Fl$, where $k_F$ is the Fermi wave vector and $l$ is the electronic mean free path. This parameter decreases with increasing disorder. The expression of the Ioffe-Regel parameter, $k_Fl=[(3\pi^2)^{2/3}\hbar]/(\rho n^{1/3}e^2)$, which can be derived from the Drude conductivity, is determined by the resistivity $\rho$ and carrier density $n$. Generally, the carrier density $n$ can be measured by Hall effect. In this work, disorder of NbN is described by superconducting transition temperature $T_c$, then we calculated \kFl according to the relation between \Tc and \kFl reported in Ref.~\cite{Pratap2012PhaseDiagram}. For Au films, bombardment of Ar$^+$ should not change the free electron density, which allowed us to calculate \kFl with a fixed density $n=5.91\times10^{22}~\text{cm}^{-3}$ \cite{Lee2016Electronic}. In the expression of \kFl for Au, $\rho$ is residual resistivity to describe the effect of disorder, which is measured in Physical Property Measurement System (PPMS) using the four-probe method. 

\section{experimental setup}

For the THG measurements, the intense broadband THz pulses were generated by tilted-pump-pulse-front optical rectification in a LiNbO$_3$ crystal, which was driven by an amplified Ti:Sapphire laser with a center wavelength of 800~nm. Two bandpass filters were inserted before the sample to generate fundamental narrow-band pump pulses and a bandpass filter with a frequency of the third-harmonic was inserted after the sample to weaken the fundamental intensity. Intense fundamental THz pulses are focused on the sample, generating pulses at the third harmonic frequency. The waveforms of the pulses are measured by electro-optic sampling using a 1-mm-thick ZnTe crystal. For the measurements without magnetic field, a helium-flow cryostat was utilized to provide a tunable temperature environment low to 5~K. The small size of the cryostat allow us to tightly focus the THz beam with a focal length of 2 inches. In measurements involving magnetic field dependence, we utilized a superconducting magnet to provide magnetic field of up to 10~T, while offering a lower temperature environment reaching as low as 2~K. In this experiment, the magnetic field is perpendicular to the sample surface, as shown in Fig.~\ref{setup}. The whole schematic diagram of the setup is shown in Ref.~\cite{Wang2024Tabletop}. Due to the large size of the magnet, only the OAP mirrors with focal lengths longer than 8 inches can be used, which makes it impossible to tightly focus the THz beam as using the cryostat. The 0.42~THz field strength in the magnet is about 14.6~kV/cm.

\begin{figure}
	\centering
	\includegraphics[width=2.6in]{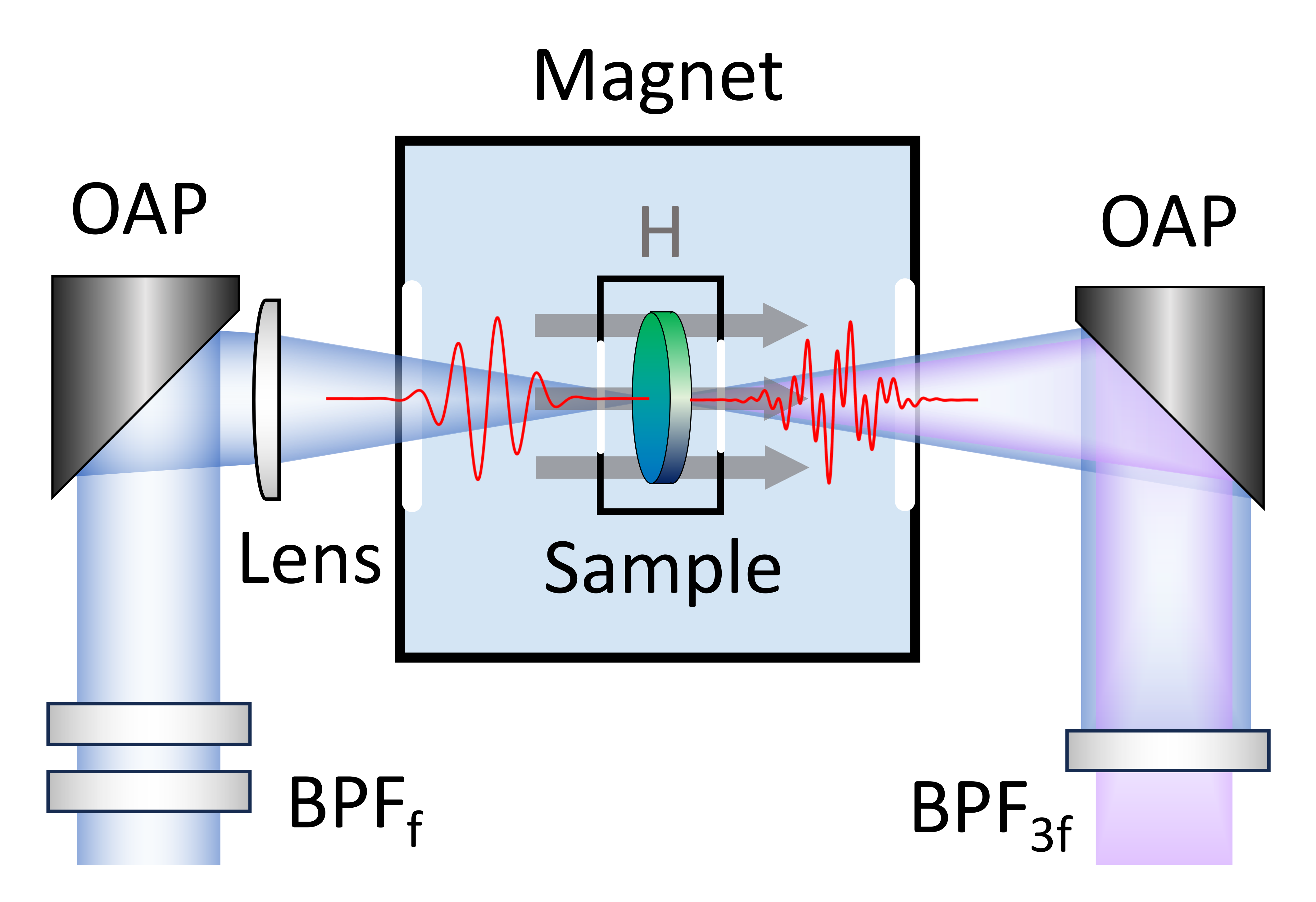}
	
	\caption{Measurement scheme of THG in the magnet.}
	
	\label{setup}
\end{figure}

%To investigate the polarization dependence, we rotated the sample instead of changing the polarization direction of THz beam, which could avoid weakening the THz intensity by non-parallel polarizers and simplify the experiment. A nonmagnetic rotational stage (MultiFields Tech) was mounted at the end of the sample rod to achieve this experimental scheme.

\section{Estimation of the electric field strength within the samples}

In Fig.~\ref{fig:NbN THG} of the main text, THG is measured under different electric field strengths of 0.42~THz pump. The samples with \Tc = 15~K, 13~K, and 5.9~K are pumped with peak field strengths of 37.8~kV/cm, 50.4~kV/cm, and 14.6~kV/cm, respectively. According to the THz time-domain spectroscopy results, the transmission coefficients at 0.42~THz in normal state (with the Fabry–Pérot effect inside the sample taken into account) are 0.034 (optical parameter from Ref.~\cite{Demsar2011EnergyGap}), 0.042 and 0.146 for the three samples, respectively. The corresponding internal field strengths are therefore estimated to be 1.29~kV/cm, 2.12~kV/cm and 2.13~kV/cm, indicating nearly identical field strengths inside the samples with $T_c=$ 13~K and 5.9~K.

\section{THG measurement result of NbN with $T_c$ = 5.9~K}

\begin{figure}
	\centering
	\includegraphics[width=6.5in]{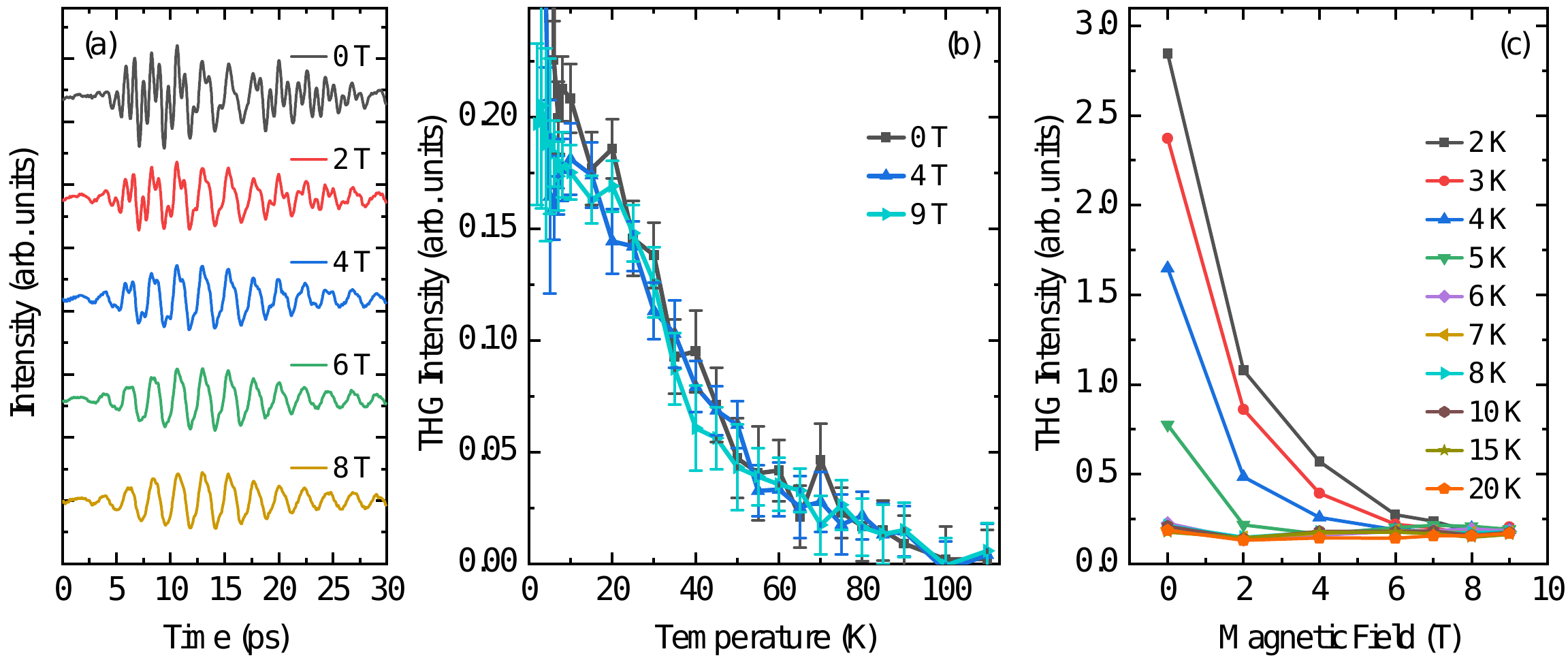}
	
	\caption{THG measurement of the disordered NbN film with \Tc = 5.9~K under magnetic field, driven by 0.42~THz. (a) Original time-domain transmitted waveforms measured at 2~K under different magnetic fields, corresponding to Fig.~\ref{fig:NbN6K}(c,d) in the main text. (b) Temperature dependence of THG intensity in normal state under different magnetic fields with error bars. (c) THG intensity as a function of magnetic field at different temperatures.}
	
	\label{NbNTHG}
\end{figure}

\begin{figure}
	\centering
	\includegraphics[width=5in]{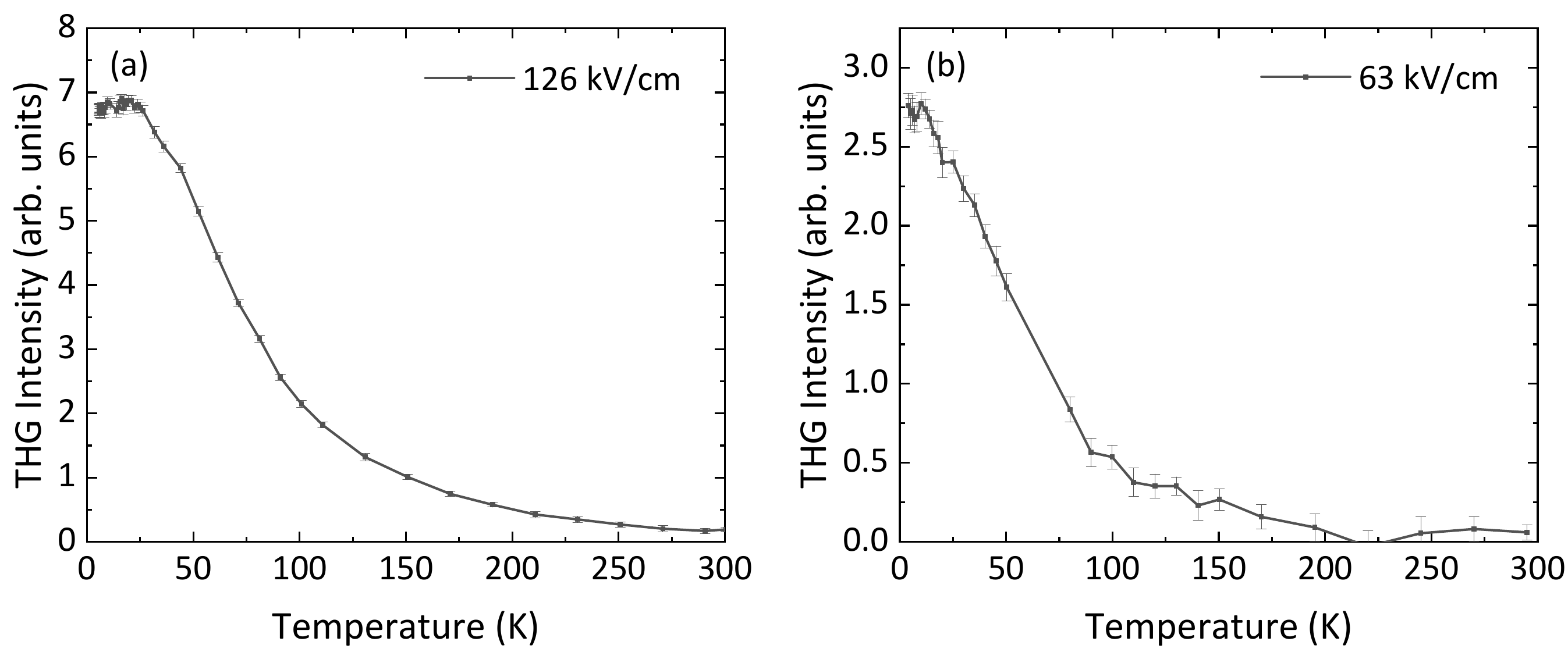}
	
	\caption{THG measurements of the disordered NbN film with \Tc = 5.9~K in cryostat, driven by 0.7~THz pulses with peak electric field strengths of (a) 126~kV/cm and (b) 63~kV/cm.}
	
	\label{0.7}
\end{figure}

\begin{figure}
	\centering
	\includegraphics[width=3in]{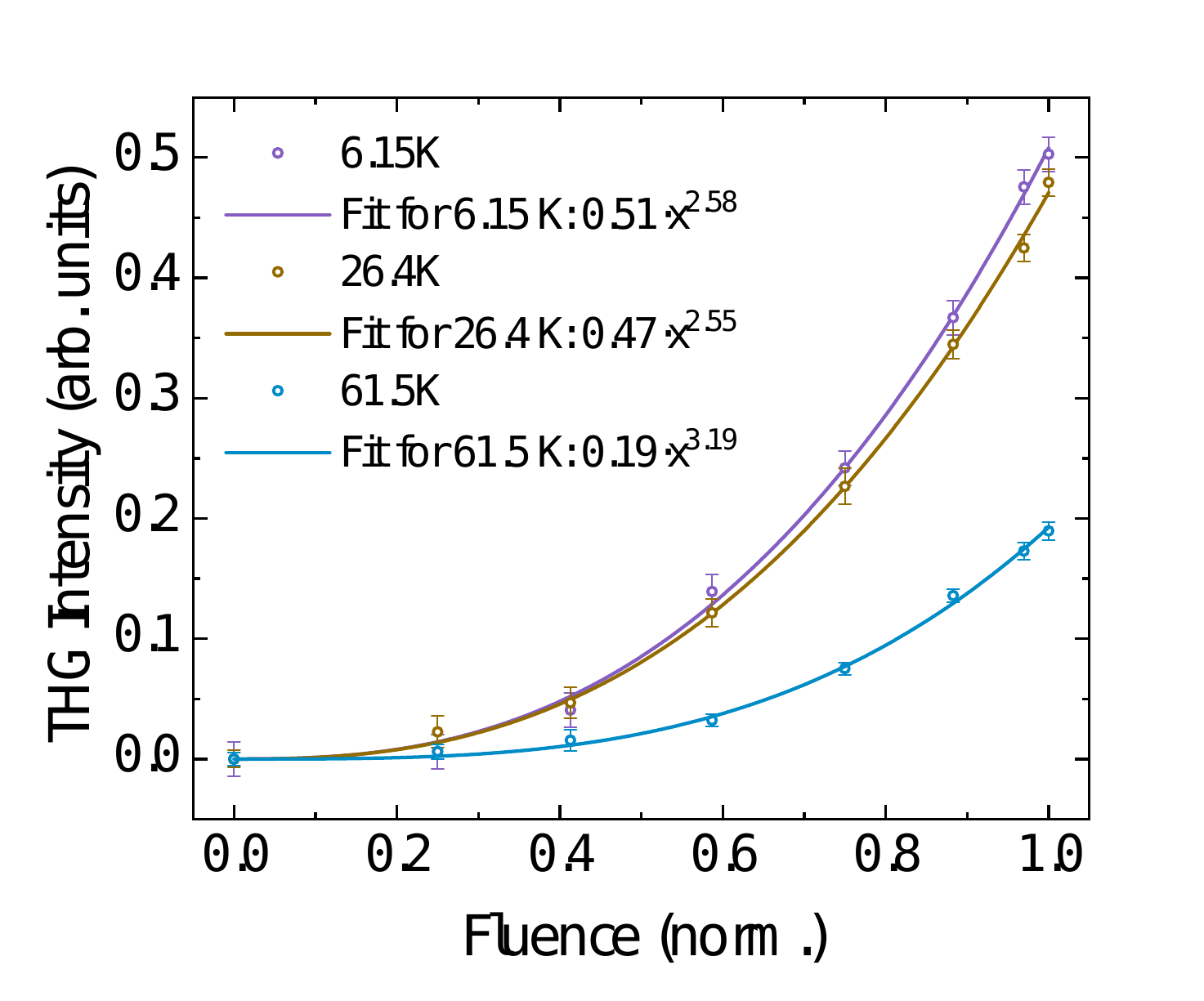}
	
	\caption{THG intensity as a function of pump electric field strength.}
	
	\label{fluence}
\end{figure}

\begin{figure}
	\centering
	\includegraphics[width=6.5in]{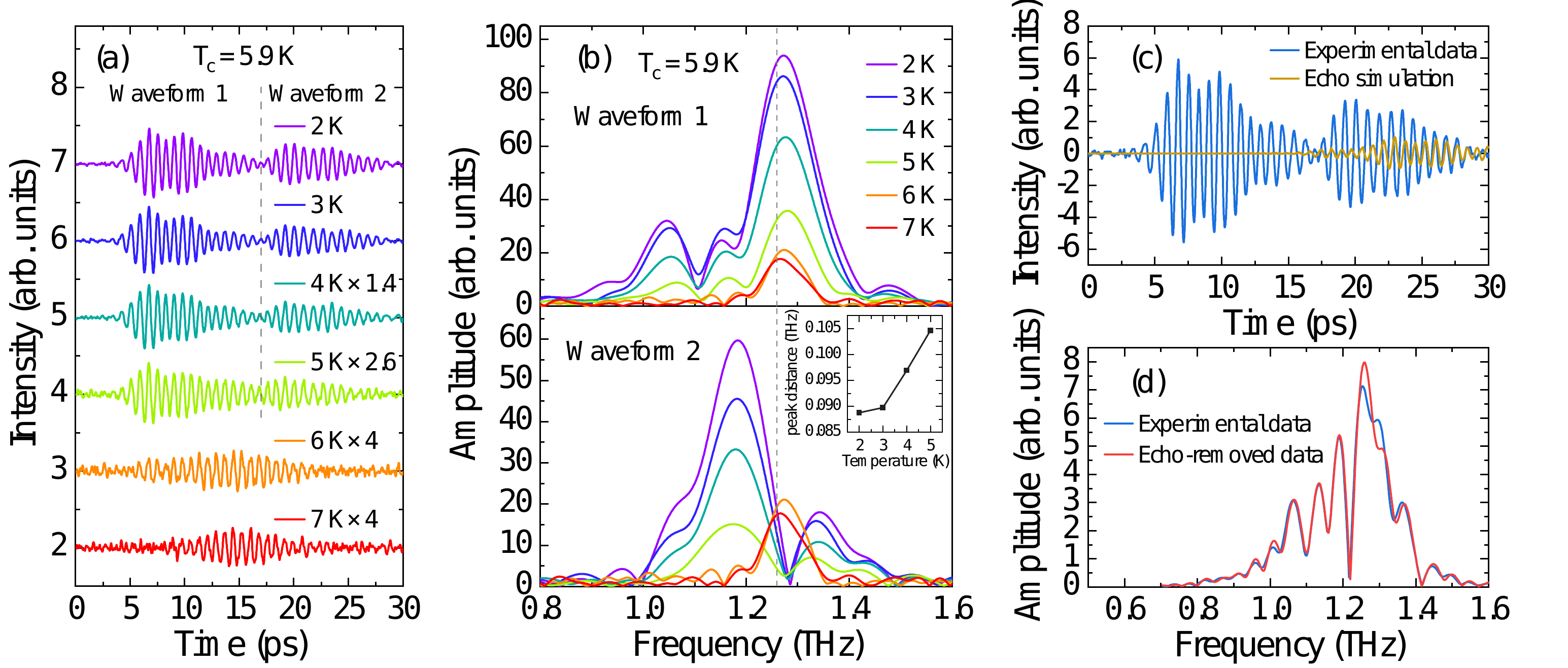}
	
	\caption{(a) 0.75~THz high-pass digital-filtered time-domain waveforms of NbN with $T_c = 5.9$~K at different temperatures without magnetic field. (b) Piecewise FFT of the THG waveforms in (a). The dashed gray line in (a) marks the cutoff point used to divide the waveforms in superconducting state. FFTs of the segments on the left and right of the cutoff line are shown in the upper and lower panels of (b), respectively. For the normal state (6~K and 7~K), FFTs of the whole waveforms are displayed in both panels. The inset in (b) shows the separation between the main peaks of the two panels. (c) THG waveform at 2~K and its echo simulation. (d) FFT spectrum of the experimental data, and the FFT spectrum of the waveform with echoes removed.}
	
	\label{piecewiseFFT}
\end{figure}

\begin{figure}
	\centering
	\includegraphics[width=6.5in]{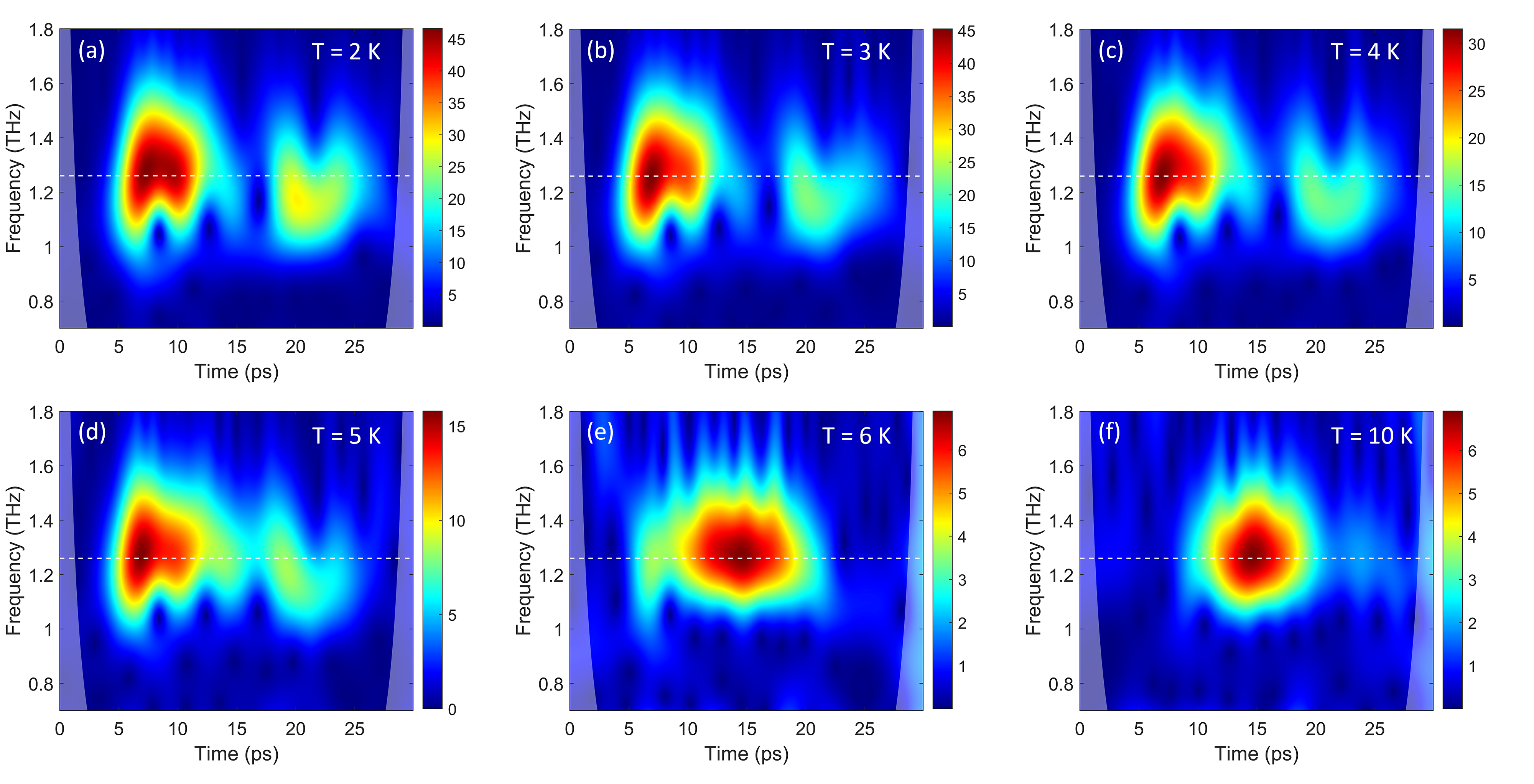}
	
	\caption{Wavelet transforms of the time-domain waveforms of NbN with $T_c$ = 5.9~K in the magnet at different temperatures. The white dashed lines represent the THG frequency at 1.26~THz.}
	
	\label{wavelet}
\end{figure}

In Fig.~\ref{NbNTHG}, we show some THG measurement results of the NbN with $T_c$ = 5.9~K in the magnet as supplement of Fig.~\ref{fig:NbN6K}. In Fig.~\ref{NbNTHG}(b), we show some temperature dependence of THG intensity above $T_c$ with error bars, corresponding to the inset in Fig.~\ref{fig:NbN6K}(e) of the main text. The error bars are defined as the noise floor of the frequency spectra. The figure shows that, within our measurement resolution, the THG intensity remains constant under different magnetic fields. Figure~\ref{NbNTHG}(c) shows the THG intensity as a function of magnetic field at different temperatures, which can show the intactness of THG in normal state under magnetic fields more clearly. In this measurement, the THG intensity disappears at about 100~K. Actually, the disappearance temperature depends on the driving field strength and signal to noise ratio. Figure~\ref{0.7} shows the temperature dependence of THG intensity driven by 0.7~THz. Under a driving field of 126~kV/cm, the THG signal remains clearly observable up to 300~K. In contrast, when the field strength is reduced to 63~kV/cm, the THG becomes indistinguishable above 200~K. 

We measure the fluence dependence of THG and fit them with power function, as shown in Fig.~\ref{fluence}. The data is measured in cryostat for better signal. The THG intensity follows $|E_{\mathrm{pump}}|^3$ dependence, confirming the signal arises from the third-order nonlinearity.

The beating feature in the time-domain waveforms can be observed clearly in Fig.~\ref{piecewiseFFT}(a) and Fig.~\ref{wavelet}. In the superconducting state, the THG waveforms are separated into two parts, each of which appears to consist of multiple components. Figures~\ref{piecewiseFFT}(b) and \ref{wavelet} indicate that the first wave packet have a peak frequency higher than 1.26~THz, and the frequency of the second wave packet is much lower than 1.26~THz. The separation of the two frequency spectra is attributed to the coupling between normal state channel and the superconducting channel, causing a energy level repulsion. With increasing temperature, the contribution from the superconducting channel gradually decreases and approaches that of the normal-state channel. This enhances the coupling strength and then increases energy level repulsion, leading to an increasing frequency separation, as shown in the inset of Fig.~\ref{piecewiseFFT}(b). When the temperature or magnetic field exceeds the critical value, only a single wave packet appears in the middle, the frequency of which is around 1.26~THz. This indicates that the anomalous features are driven by superconductivity.

\begin{table}[htbp]
	\centering
	\caption{Estimated time delays of echoes from optical elements.}
	\begin{tabular}{l@{\hspace{0.5cm}}c@{\hspace{0.5cm}}c}
		\toprule
		Optical Element & Thickness (mm) & Time Delay (ps) \\
		\midrule
		Substrate (MgO) & 0.5 & 10.8 \\
		Inner magnet window (diamond) & 1.0 & 15.9 \\
		Outer magnet window (z-cut Quartz) & 3.0 & 40 \\
		Electro-optic sampling crystal (ZnTe) & 1.0 & 21 \\
		\bottomrule
	\end{tabular}
	\label{tab:echo delay}
\end{table}

We conclude that the observed beating feature is not caused by echoes from other optical elements in the beam path, based on the following reasons. We calculate the time delays induced by echoes from potential optical elements, as shown in Table~\ref{tab:echo delay}. The time delay of the second wave packet is about 13~ps, which does not correspond to any of the optical elements considered. Furthermore, the echoes do not change the frequency spectra range. For the sample with $T_c = 5.9$~K, as shown in Fig.~\ref{piecewiseFFT}(b) and Fig.~\ref{wavelet}, the evidently different spectra rule out the possibility of echoes. Additionally, the echoes are not influenced by the sample state, whether it is in the superconducting or normal state. The intensity of the second wave packet reaches 58\% of that of the first one. If it were caused by echoes, a similar feature with the same intensity ratio should also be observed at the same time delay ($\sim$13~ps) in the waveform measured at 7~K. Our noise level is low enough to resolve signals with intensities greater than half of the main peak. 

We simulate the echoes of the THG waveform at 2 K generated by the substrate and the diamond window, as shown in Fig.~\ref{piecewiseFFT}(c). The intensity of the echoes is much weaker than that of the second wave packet. As shown in Fig.~\ref{piecewiseFFT}(d), removing the echoes from the experimental waveform only induces a slight change in the FFT spectrum. The simulation indicates that the influence of echoes is negligible.

\begin{figure}[h]
	\centering
	\includegraphics[width=2.5in]{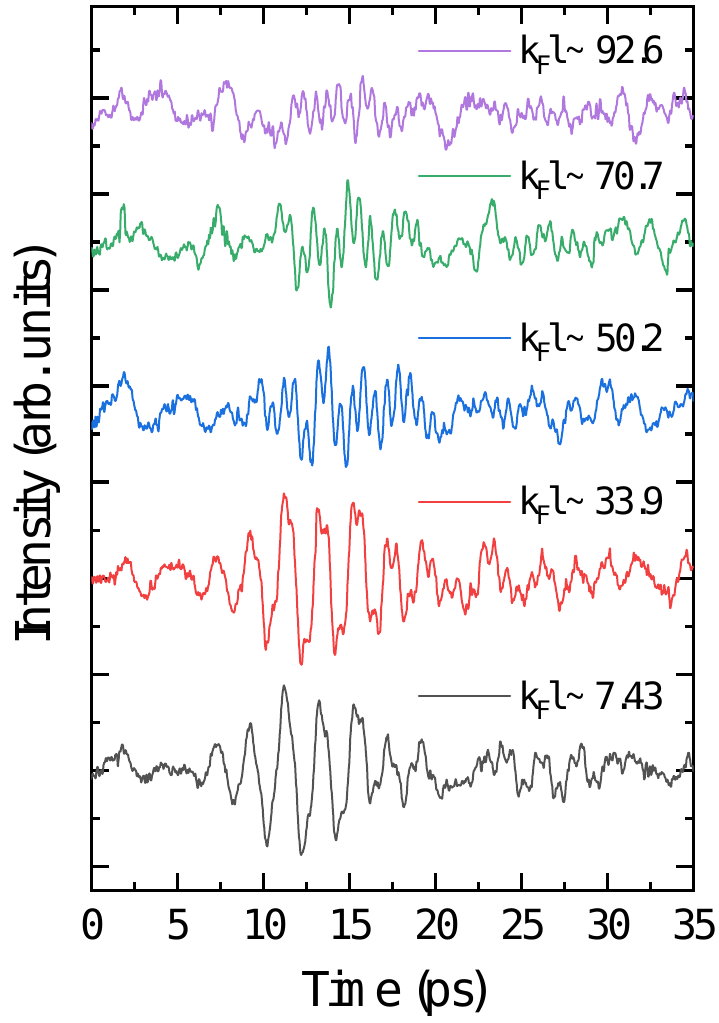}	
	\caption{Time-domain waveforms in THG measurement for Au with different \kFl, measured at 5~K.}
	
	\label{AuTHG}
\end{figure}

\section{THG measurement result of Au}

Corresponding to the digital-filtered waveforms in Fig.~\ref{fig:Au THG}(a) of main text, the original measured waveforms are shown in Fig.~\ref{AuTHG}.

\section{calculation of nonlinear coefficient $\chi^{(3)}$}

The calculation of nonlinear coefficient is given by $\chi^{(3)} \propto E_{3\omega}^{\mathrm{in}}/(E_{\omega}^{\mathrm{in}})^3$, where $E_{3\omega}^{\mathrm{in}}$ and $E_{\omega}^{\mathrm{in}}$ represent the electric field strengths inside the sample. In experiment, we can only measure the field strength outside the sample ($E_{3\omega/\omega}^{\mathrm{out}}$). The calculation can be written as $|\chi^{(3)}| \propto |E_{3\omega}^{\mathrm{out}}/\tilde{t}_{3\omega}|/|E_{\omega}^{\mathrm{out}}/\tilde{t}_{\omega}|^3$, where $\tilde{t}$ is the transmission coefficient from sample to substrate. In THz range, the refractive index of the metal is significantly higher than that of the substrate. According to the Fresnel equation, $|\tilde{t}| = |2\tilde{n}_{Au}/(\tilde{n}_{\mathrm{Au}}+\tilde{n}_{\mathrm{MgO}})| \approx 2$. 
The correction factor $|\tilde{t}_{\omega}^{3}/\tilde{t}_{3\omega}|$ is nearly constant in our experiment, allowing us to calculate the nonlinear coefficient through $|\chi^{(3)}| \propto |E_{3\omega}^{\mathrm{out}}/(E_{\omega}^{\mathrm{out}})^3| \propto I_{3\omega}/I_{\omega}^3$, where $I_{3\omega/\omega}$ is the integral of $3\omega/\omega$ peak in the frequency spectrum. $I_{\omega}$ is measured without $3\omega$-bandpass filter for better signal.

\section{Larkin-Ovchinnikov model}
The key parameters of the LO model are superconducting gap $\Delta$ and the depairing strength $\eta$, they couple in the Usadel equation
\begin{equation}
	iE \sin\theta + \Delta \cos\theta - \eta \Delta \sin\theta \cos\theta = 0,
\end{equation}
where $E$ is the energy relative to Fermi level, $\theta$ is the pairing angle, and $\sin\theta$ and $\cos\theta$ are the quasiclassical, disorder-averaged Green’s functions \cite{Klapwijk2012LO}.

The real part of optical conductivity in superconducting state can be calculated by a generalized Mattis-Bardeen equation \cite{Nam1967LOfit}
\begin{align}
	\frac{\sigma_{1s}}{\sigma_{1n}} 
	= & \frac{2}{\hbar \omega} \int_{E_g}^{\infty} \left[ f(E) - f(E + \hbar \omega) \right] g_1(E,E+\hbar \omega) \, \mathrm{d}E \notag \\
	& + \frac{1}{\hbar \omega} \int_{E_g - \hbar \omega}^{-E_g} \left[ 1 - 2f(E + \hbar \omega) \right] g_1(E,E+\hbar \omega) \, \mathrm{d}E,
\end{align}
where $f(E)$ is the Fermi-Dirac distribution function. Energy gap is renormalized to $E_g(\eta) = (1 - \eta^{2/3})^{3/2} \Delta$. The generalized coherence factor is given by
\begin{equation}
	g_1(E,E+\hbar\omega)=\mathrm{Re}[\cos\theta(E)]\mathrm{Re}[\cos\theta(E+\hbar\omega)]+\mathrm{Im}[\sin\theta(E)\mathrm{Im}[\sin\theta(E+\hbar\omega)]].
\end{equation}
When $\eta=0$, the model recovers the traditional Mattis-Bardeen form. 

\end{document}